\documentclass[superscriptaddress,aps,preprint,longbibliography]{revtex4-2}
\usepackage{graphicx} 
\usepackage{amsmath}
\usepackage{amssymb}
\usepackage[utf8]{inputenc}
\usepackage{booktabs}
\usepackage{enumitem}
\usepackage{xcolor}
\usepackage{lineno}

\graphicspath{{./pictures/}}
\begin{document}
\title{Predicting tipping points: The many shades of non-equilibrium and catch-22 of early-warning}
\author{Johannes Lohmann}
\affiliation{Physics of Ice, Climate and Earth, Niels Bohr Institute, University of Copenhagen, Denmark}

\begin{abstract}
The potential of crossing climate tipping points has reached the attention of many researchers and the general public. On the one hand, the basis for this concern is strengthening, with simulations showing abrupt transitions that may occur even for moderate greenhouse gas emission scenarios. On the other hand, our understanding of what constitutes such transitions mathematically is becoming more nuanced. This leads to challenges for the fidelity of early-warning signals (EWS), which accompany bifurcations in systems that closely track a slowly changing steady state. Different kinds of out-of-equilibrium dynamics in a rapidly changing climate, as well as chaos, blur the notion of a critical tipping threshold and imply limits in predictability.
Using EWS to detect loss of local stability from data requires careful evaluation in the high-dimensional system, and may have limited predictive power in a highly multistable climate, where there is more than a single, well-known alternative state. This calls for new methods using non-equilibrium statistical mechanics and dynamical systems to efficiently probe the global stability properties of climate models and observations.
\end{abstract}

\maketitle

\newpage

\section{Introduction}

It is of great current interest to know whether singular, irreversible critical transitions can be expected in large-scale sub-systems of the Earth, and whether these can be anticipated from observational data before a point of no return is reached. It has been argued on physical grounds that such Earth system tipping points should exist, and on mathematical grounds that they can be predicted, even if they cannot be accurately simulated.


For the remained of the paper, the terms ``tipping point'' (TP) and ``tipping'' are used as follows. A TP (in the narrow sense) is a transition from one stable steady state to another, the two of which coexist for at least some interval of a control parameter or boundary conditions. These transitions happen when:

1. The system crosses a bifurcation (more generally a boundary crisis of an attractor), where one of the two steady states loses stability upon a time-varying change of a parameter that reaches a critical threshold.

2. The system’s trajectory crosses a basin boundary due to, a) external, stochastic forcing (at fixed parameters), or b) a time-varying parameter change while the (autonomous) system does not cross bifurcation.

These {\it tipping mechanisms} correspond to the widely used terminology of bifurcation-, rate-, and noise-induced tipping \cite{ASH12}.
TPs in the narrow sense are said to be {\it irreversible} transitions, because - owing to multistability over a finite parameter interval - when the alternative state is reached it requires a finite parameter change (or substantial noise forcing) to return to the previous state. The required change can be larger than the one that led to the transition.
In a widened sense, a TP can also refer to transitions due to excitability, where a “quasithreshold” \cite{WIE23} in phase space can be crossed (due the above mechanisms 2a and 2b) that leads to a large, temporary excursion away from - and then returning to - a single steady state. However, it may not be easy to define in practise at what point one can speak of a sufficiently large “excursion”.

Besides some paleoclimate evidence \cite{BRO21}, the physical justification for TPs comes from {\it positive feedback} processes \cite{LOR25}, such as the salt-advection feedback in the Atlantic ocean, the melt-elevation feedback of the polar ice sheets, and the forest-rainfall feedback in the Amazon. Together with a globally stabilizing negative feedback and non-linearity, a positive feedback can lead to multistability, i.e., the existence of an alternative stable state (in addition to the ``present-day'' state) that may be reached via a TP.

\begin{figure}
\includegraphics[width=0.49\textwidth]{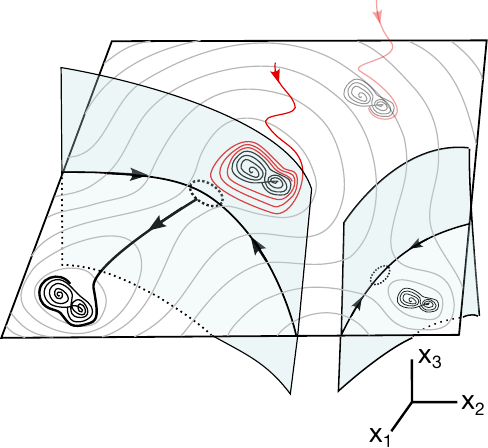}
\caption{\label{fig:csd}
Illustration of critical slowing down (CSD) for a system with co-existing chaotic attractors. The top right shows the present-day attractor far from a bifurcation, where a perturbation away from the $(x_1 - x_2)$-plane leads to a quick relaxation in all degrees of freedom (red trajectory). As the system approaches the TP (middle of picture), the attractor approaches the basin boundary (vertical, curved surface) and its embedded saddle (dotted ellipse). CSD is experienced, where the relaxation after perturbation gets very slow in the $(x_1 - x_2)$-plane, and in particular towards the saddle.
In the front left is the (undesired) attractor that would be reached after tipping, and in the front right is another co-existing attractor and its basin boundary.
From CSD alone it is not obvious which of the two alternative states would be reached. A slice across the quasi-potential, as would be obtained under stochastic forcing \cite{Graham1991}, is given by the contours in the $(x_1 - x_2)$-plane.
}
\end{figure}

The mathematical argument for TP predictability relies on the above mechanism 1, i.e., TPs as a result of generic bifurcations of a dynamical system. The destabilization of a present-day state is then preceded by critical slowing down (CSD) (see Fig.~\ref{fig:csd}), which under the action of stochastic environmental perturbations may be measured by statistical early-warning signals (EWS) that have been considered “universal” precursors of a TP \cite{SCH09}.
This idea has been appealing for a long time, but it is now being scrutinized in earnest as the focus has evolved from paleoclimate observations \cite{DAK08} to near-future scenarios, and as more sophisticated models simulate what appear to be TPs. Several recent studies propose EWS in historic observations as evidence of approaching TPs in the Atlantic meridional overturning circulation (AMOC) \cite{BOE21a,MIC22,DIT23}, the Greenland ice sheet \cite{BOE21}, the Amazon rainforest \cite{BOU22}, and the South American monsoon \cite{BOC23}. The studies assume that the system will undergo a saddle-node bifurcation.

This paper gives a perspective on whether it is really possible, given our current knowledge, to anticipate any large-scale, irreversible changes in climate sub-systems such as the ones mentioned above. In such complex systems, i.e., not homogeneous, controlled systems undergoing, say, a well-known equilibrium phase transition, a variety of complicating factors are discussed that call into question a) our ability to extract a CSD signal from observations, and b) the predictive power of CSD regarding what happens next to the system. Further, as opposed to other difficult prediction problems such as earthquakes or market crashes, we cannot easily validate the methodology on previously observed events. Climate TPs due to anthropogenic change are a ``zero-shot'' prediction problem, which suggests that a very strong theoretical basis should be in place first.

As motivating example, consider a possible collapse of the AMOC. Here, the existence of a positive feedback (salt-advection) is plausible \cite{LIU17,DIJ26} and simulations with sophisticated models show a transition to a stable collapsed circulation using (not necessarily realistic) meltwater forcing as control parameter \cite{VWE24}. Using (more realistic) CO$_2$ forcing, some climate models of the current generation also show that collapses can occur within the next two centuries \cite{ROM23,DRI25}.
On the other hand, it is not clear whether such simulated collapses are true, singular transitions between (two!) alternative stable states or a transient response \cite{CUR24} (which could include excitability as tipping in the widened sense), and whether they tell the entire story of the turbulent ocean and coupled Earth system:
The response of the climate system to CO$_2$ forcing does not only affect the AMOC's positive feedback, but many other competing processes, some of which may not be properly represented in model simulations. This includes stabilizing influences of Antarctic meltwater \cite{SIN25} and Indian ocean warming \cite{HU19}, an adaptive response of the ocean by ``Atlantification'' of Arctic \cite{ART25}, and a slowdown of Greenland melt as result of an AMOC slowdown \cite{POP25}.
Moreover, a collapse of the AMOC may occur as a sequence of bifurcations in the spatio-temporal pattern of the circulation \cite{RAH95,LOH24,THI26}.
Finally, the changes in CO$_2$ forcing are fast compared to the slow deep ocean time scales, and transitions seen in models could be non-autonomous effects such as excitable transients or rate-induced tipping \cite{STO97,LOH21,VWE26b}, and not bifurcation crossings.
These considerations imply various challenges for the prediction of TPs in the Earth system, which will be explained in Sec.~\ref{sec:challenges} before discussing possible directions for future research in Sec.~\ref{sec:directions}.




\section{Challenges for anticipating Earth system tipping points}
\label{sec:challenges}

\begin{figure}
\includegraphics[width=0.99\textwidth]{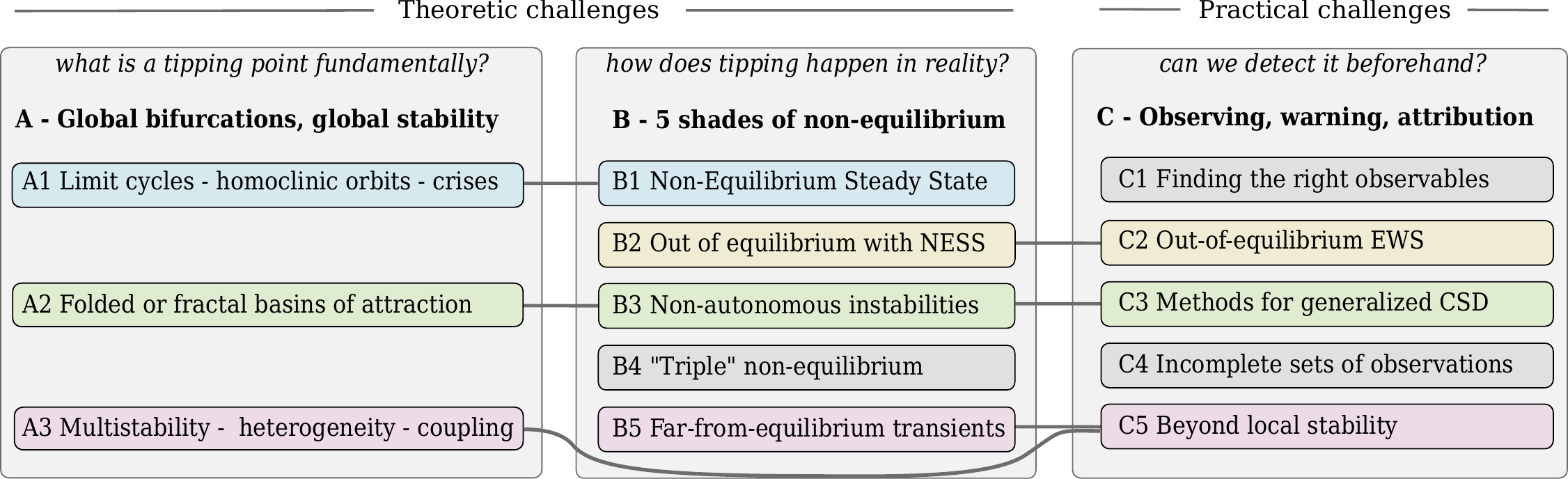}
\caption{\label{fig:schematic}
Challenges for our understanding and the prediction of tipping points in the Earth system. Items that are related across the different theoretic and practical dimensions are connected and shaded in the same color.
}
\end{figure}

It is hard to infer from realistic models what the mathematical nature of a TP is, since long equilibrium simulations are needed. Conversely, these might also not tell the full picture in the out-of-equilibrium climate, which depends on the global stability landscape. As quantitative predictions across models remain uncertain, generic approaches based on dynamical systems and statistical mechanics are appealing, but it needs to be clarified how they apply to the Earth system. Figure~\ref{fig:schematic} gives an overview of different dimensions in which our understanding of Earth system TPs and the possibility for early-warning needs to be improved. This pertains to the mathematical nature of a TP (A1-A3), the way TPs happen in reality (B1-B5), and how in practise they can be anticipated from data (C1-C5). The following three sections explain this in more detail.


\subsection{Global bifurcations and global stability}
\label{sec:global}

\begin{figure}
\includegraphics[width=0.75\textwidth]{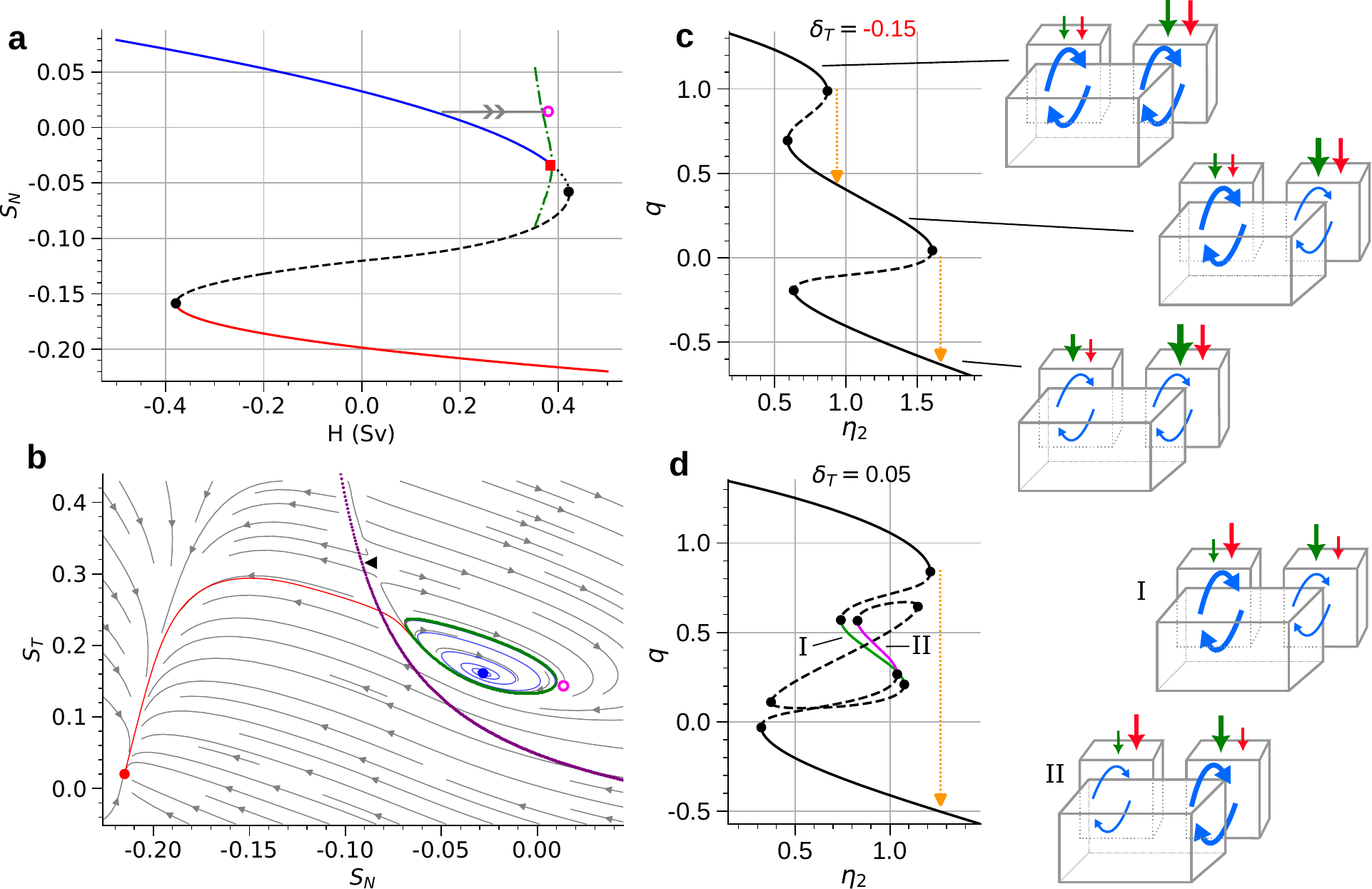}
\caption{\label{fig:bifurcations}
{\bf a,b} Tipping in a conceptual box model of the AMOC \cite{WOO19,ALK19}. Model equations and parameters can be found in \cite{ALK19}.
{\bf a} Bifurcation diagram of the variable $S_N$ (proportional to the AMOC strength) with freshwater forcing $H$ as control parameter. Solid lines are stable fixed points, the dashed line is a saddle, and the dash-dotted line is an unstable periodic orbit. The filled dots are fold bifurcations, and the square is a subcritical Hopf bifurcation.
{\bf b} Phase portrait of the system at $H=0.37$, i.e., shortly before the subcritical Hopf. The blue line is the unstable periodic orbit, which forms the basin boundary. The basin boundary (purple line) and stable fixed point (open pink circle) at $H=0.15$ are also shown.
{\bf c,d} Another conceptual AMOC model (equations and parameters as in \cite{LOH26}) with spatially heterogeneous freshwater and temperature forcing (indicated by the size of the green and red arrows) in two northern boxes. Bifurcation diagrams with mean freshwater forcing $\eta_2$ as control parameter are given for two model configurations, one with consistent ({\bf c}) and one with opposing heterogeneity in freshwater and temperature forcing ({\bf d}). The qualitative state of the overturning circulation in the different stable states is indicated by the blue arrows.
}
\end{figure}

The possible existence of bifurcations other than the saddle-node, as well as a more complex global stability landscape, makes it harder to anticipate a TP.
On the one hand, EWS can appear for non-catastrophic bifurcations or transitions without bifurcation (but high parameter sensitivity) \cite{KEF13,RIE25}.
On the other hand, the saddle-node bifurcation may be replaced by global bifurcations such as the sequence of homoclinic and subcritical Hopf bifurcations. This is illustrated for a conceptual AMOC model in Fig.~\ref{fig:bifurcations}a,b. Before the present-day AMOC fixed point (blue line) can collide with the saddle, an unstable periodic orbit (green dot-dashed line) appears at a homoclinic bifurcation and shrinks at a subcritical Hopf bifurcation onto the present-day state, which becomes unstable. This has been observed not only in conceptual models describing the AMOC's salt-advection feedback \cite{SCO99,TIT02,TIT02a,ALK19}, but also in Welander's model of the convective feedback \cite{WEL82,CES96,ABS04} (a different mechanism for AMOC destabilization),
and it can exist in other contexts where a bistable element is coupled via a negative feedback to a slower-timescale system, for instance the ice sheet coupled to a bedrock that is evolving dynamically due to viscoelastic deformations of the mantle (not shown here).
Though this sequence of bifurcations depends on model parameters, it may be a viable tipping mechanism just like the saddle-node bifurcation. There are a few consequences for early-warning: The Hopf bifurcation upends the square-root scaling of the equilibrium with the control parameter before a SN bifurcation, which is sometimes evoked as useful hallmark of tipping \cite{DIT23,VWE25c}. There is CSD, but the relaxation rate decreases linearly with the control parameter, and not as a square root like for the SN.

Perhaps most interestingly, as the unstable periodic orbit is created at the (global) homoclinic bifurcation, the basin of attraction becomes finite (Fig.~\ref{fig:bifurcations}b). The trajectory at one parameter value can thus fall outside of the basin of the desired base state at another parameter value (open circles in Fig.~\ref{fig:bifurcations}a,b). The system can then tip without crossing the bifurcation where the base state loses stability (so-called rate-induced tipping, see next section).

As more degrees of freedom (d.o.f.) and processes are added, the dynamics likely becomes chaotic and the bifurcations more complicated. The SN bifurcation is replaced by global bifurcations of limit cycles or crises of chaotic attractors.
While we would still expect CSD due to a decrease in stability of the attractor \cite{TAN18,LUC24}, it may not be detectable in the same way as for fixed points where increasing noise-driven fluctuations around a mean value are harnessed. The existence and unknown change of oscillatory modes embedded in the chaotic dynamics interferes with the usual statistical EWS \cite{LOH24}. Instead, one needs special techniques to first define suitable observables (see Sec.~\ref{sec:ews}).


Another aspect is the existence of more than one alternative state, as indicated in the bottom right of Fig.~\ref{fig:csd}. This has two consequences. First, even if there is a way to unmistakenly detect CSD, it does not tell what state is reached after tipping unless one knows what kind of fluctuations distinguish the tipping to one state from tipping to another. But this level of detailed system knowledge conflicts with the idea that EWS should be generic. Second, the present-day state may also be bounded by a bifurcation (and TP to a different state) in the past. If there is both a bifurcation for increasing (future, w.l.o.g.) and decreasing control parameter, both would be associated with CSD, and hence we cannot be sure from which point in time (i.e., value of control parameter) EWS start to behave monotonically (see \cite{LOH24,LOH26}).

Reasons for additional coexisting alternative states (beyond bistability) include spatial heterogeneity \cite{BAS22,NEF23}, additional non-linear feedback processes, and coupling to other multistable subsystems. 
A simple example is shown in Fig.~\ref{fig:bifurcations}c,d, where a model of the AMOC divides the Atlantic into one box representing the tropical/mid-latitude ocean and two boxes representing
the western and eastern parts of the high-latitude North Atlantic (in particular, the Irminger-Labrador and Nordic seas, respectively).
The two northern boxes are assumed identical, but the symmetry can be broken by choosing a stronger temperature and freshwater forcing for one box (Fig.~\ref{fig:bifurcations}c). This leads to an intermediate TP where the circulation to only one of the boxes stops, i.e., a partial AMOC collapse. Another symmetry breaking occurs when the forcing heterogeneity is counteracting, such that one box has stronger temperature, but weaker freshwater forcing (Fig.~\ref{fig:bifurcations}d). In this case, there are two partially collapsed states.


\begin{figure}
\includegraphics[width=0.49\textwidth]{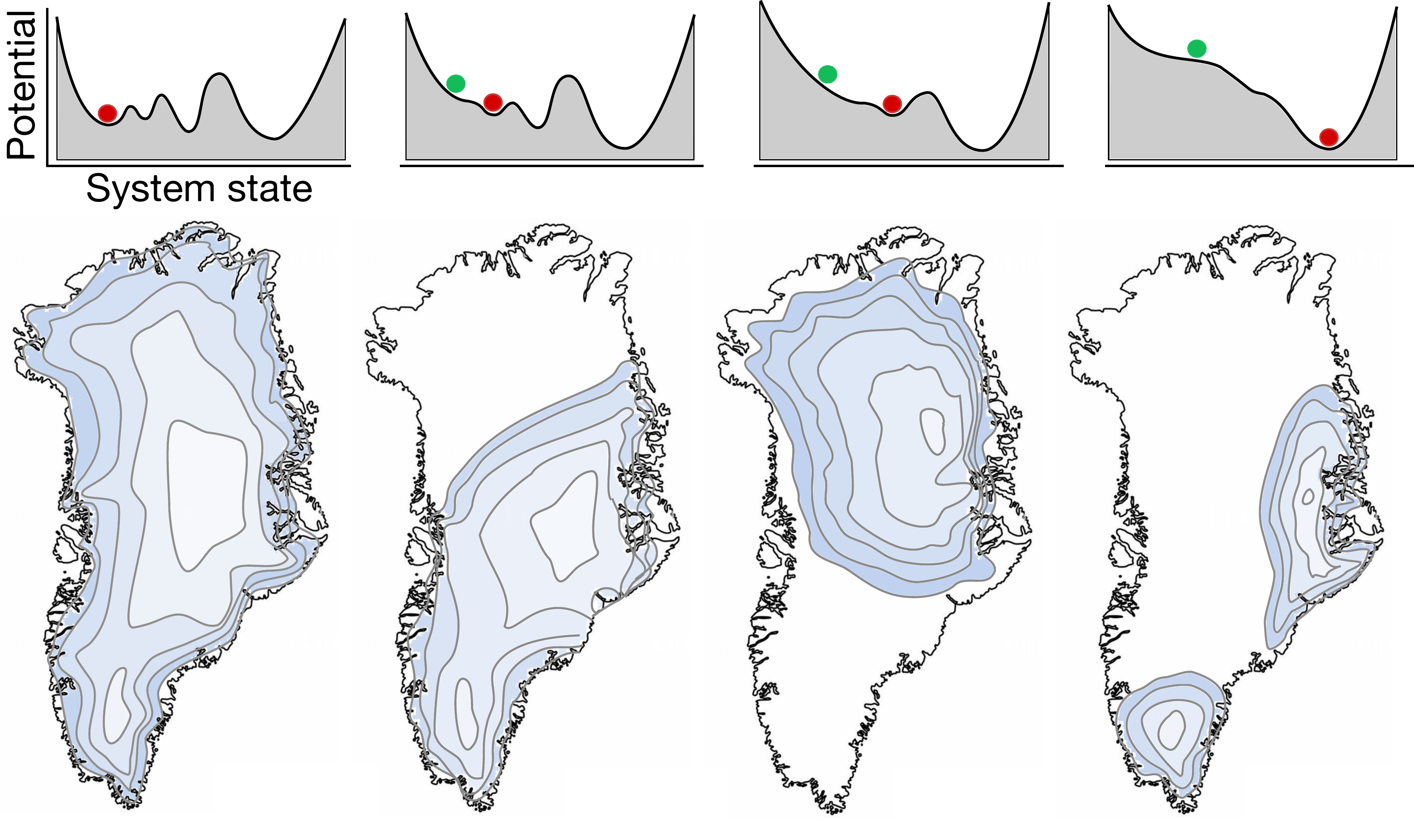}
\caption{\label{fig:greenland}
Illustration of possible co-existing steady-state configurations of the Greenland ice sheet, based on simulations of ice sheet multistability in \cite{ROB12,HOE23,AND26}. In addition to a fully collapsed ice sheet, stable semi-collapsed states with a remaining northern \cite{HOE23} or southern \cite{ROB12,AND26} ice sheet have been shown.
}
\end{figure}

A similar situation of heterogeneous, counteracting atmospheric forcing exists for the Greenland ice sheet, since southern Greenland is closer to the melting point but at the same time receives far more snowfall. Accordingly, four alternative stable states have been shown in ice sheet simulations (Fig.~\ref{fig:greenland}), including intermediate states where either the southern or northern part has collapsed \cite{ROB12,HOE23,AND26}. For Antarctica, a potential ice sheet disintegration is likely also a fragmented, step-wise process due to its even larger size and the accordingly varied boundary conditions of its marine-based outlet glaciers, leading to intermediate TPs due to bifurcations in different regions at different forcing levels \cite{WIN26,SWI26}.

It has also been suggested that large-scale ecosystems, of which the Amazon rainforest is a prominent candidate to undergo tipping, do not necessarily collapse uniformly to an undesired state (savannah or desert) in one single bifurcation \cite{RIE21}. The collapse may instead happen by pattern-formation to adapt to decreasing water resources, or in case of large-scale heterogeneity by coexistence states with broken symmetry \cite{RIE21}.
A variety of possible spatial patterns of the ocean circulation and convection was also noted \cite{RAH95,LOH24,THI26}, which can lead to sequences of subtle bifurcations that obscure a targeted prediction of an AMOC collapse by EWS.
Finally, the different suspected multistable climate sub-systems do not exist in isolation and largely react to the same control parameter (changing greenhouse gas concentrations and corresponding radiative forcing). The nature and magnitude of the coupling between sub-subsystems is still highly uncertain, but in principle it can induce a high degree of multistability in the coupled system \cite{DEK18,LOH21b}.

\subsection{Five shades of non-equilibrium}

The second column in Fig.~\ref{fig:schematic} indicates that complications for prediction of TP arise because the Earth system is out of equilibrium in several ways.
The climate can be described as a non-equilibrium thermodynamic system due to the spatially inhomogeneous insolation (B1). It thus tends towards a non-equilibrium steady state (NESS) \cite{LUC10,GHI20}. From a statistical mechanics perspective, the dynamics in this situation break detailed balance and are non-reversible.
This makes it a non-gradient dynamical system, which also means it can display oscillatory dynamics. This affects EWS, since a) fluctuation dynamics are not time-reversed paths of deterministic relaxation dynamics - i.e., the largest stochastic fluctuations may not be seen in the same d.o.f. as the slowest (critical) relaxation modes - and b) since oscillatory modes further complicate the extraction and interpretation of statistical EWS \cite{LOH24,LOH25}.

\begin{figure}
\includegraphics[width=0.92\textwidth]{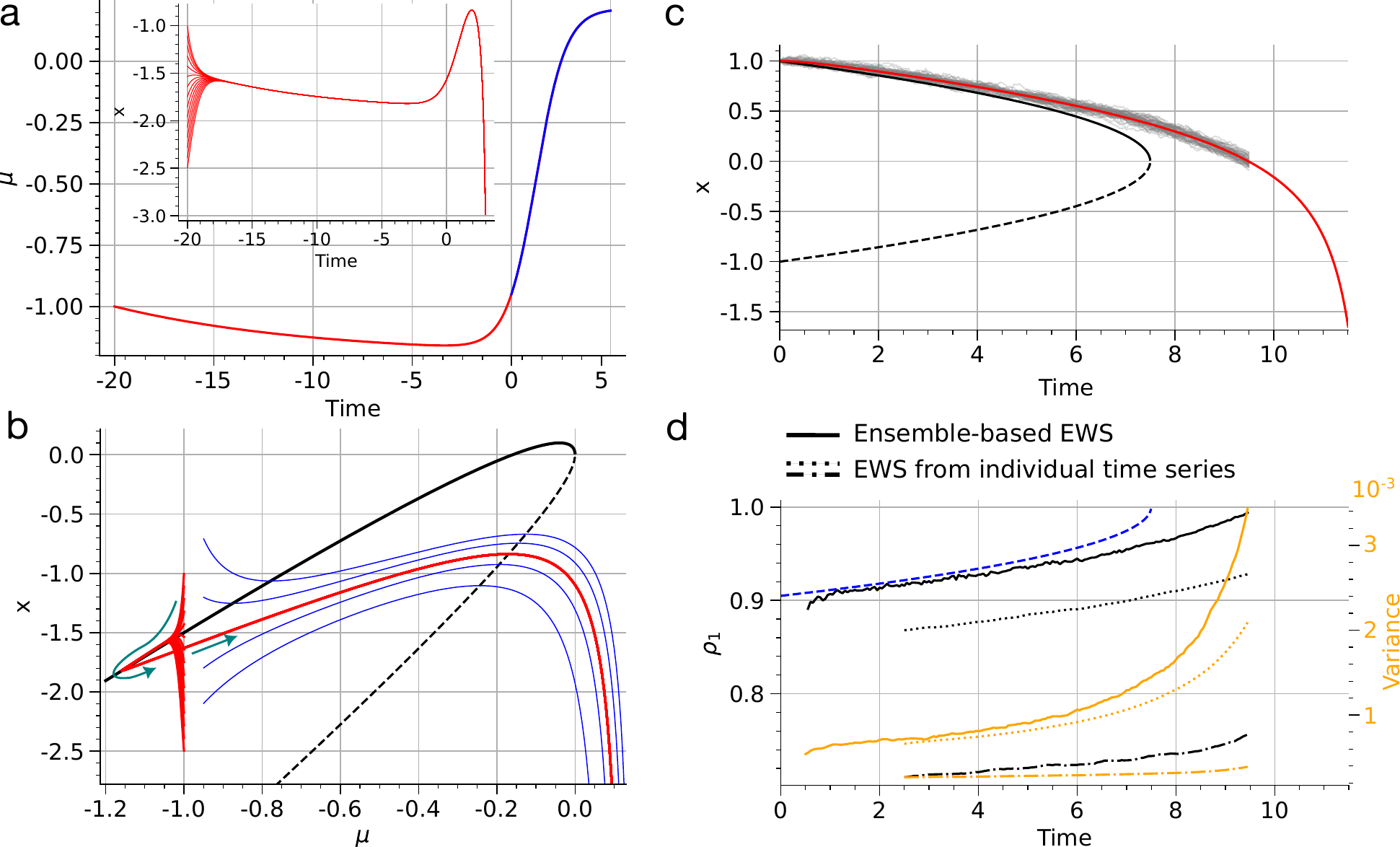}
\caption{\label{fig:nonequilibrium}
{\bf a,b} Different kinds of non-equilibrium in the 'tilted' saddle-node system $\dot{x} = -(x - \tau\mu)^2 -\mu$ with $\tau = 2.5$. {\bf a} $\mu$ is changed non-linearly, similar to the 'Hockey stick' graph of global temperature over the past millenium \cite{MAN99}. Inset: system evolution from a range of initial conditions. The trajectories converge and form the pullback attractor.
{\bf b} Stable and unstable fixed points in the autonomous system (black solid and dashed lines, respectively). The trajectories forming the pullback attractor are in red. In blue are four additional simulations under the same forcing scenario, but started only at time $t=0$.
{\bf c,d} EWS in the noisy saddle-node normal form system $dX_t = -(X_t ^2  + \mu)dt + \sigma dW_t$ with Wiener process $W_t$, under a linear increase of $\mu$ from -1 to 1 in 15 time units. {\bf c} Time varying stable (solid black line) and unstable (dashed) fixed point, as well as deterministic solution (red) and 100 simulations with noise (gray, $\sigma=0.05$). {\bf d} Ensemble variance and lag-1 autocorrelation $\rho_1$ for $N=10000$ realizations (solid lines). Dotted lines: same quantities estimated for each time series in a sliding window of width 2 time units, after subtracting the deterministic solution, and then averaged over all realizations. Dot-dashed lines: Same, but estimated after linear detrending in each window. The dashed blue line is the analytical autocorrelation in the quasi-stationary system.
}
\end{figure}

The boundary conditions that determine the NESS have changed throughout Earth's history. Even when viewed in its entirety, Earth experiences varying conditions, e.g., due to astronomical changes (time scale of thousands of years). The anthropogenic emission of greenhouse gases since the industrial period is also an external forcing, but it constitutes a monotonic change that is fast or comparable to time scales of most parts of the Earth system, perhaps with the exception of the atmosphere. This means that the system likely cannot track the changing NESS closely. Instead, it evolves on a pullback attractor \cite{CHE11}, which could be quite far from the NESS that would be attained if the boundary conditions would be held constant from a given point in time. In that sense it is in 'double' non-equilibrium - it is out-of-equilibrium with the NESS (B2).
This is illustrated in Fig.~\ref{fig:nonequilibrium}a,b, where a time-varying forcing is used that first evolves slowly (pre-industrial Holocene climate) and features a pullback attractor (panel b red line) close to the NESS (panel b solid black line), but then accelerates (industrial climate), resulting in a pullback attractor far from the NESS.

Non-autonomous instabilities, also known as rate-induced tipping \cite{ASH12,RIT23,WIE23}, are another level of non-equilibrium dynamics (B3). In some systems, the failure to closely track the changing NESS can lead to tipping before a bifurcation where the NESS loses stability. This happens in situations when during a parameter shift a trajectory crosses a basin boundary of the fixed-parameter system at a future time \cite{WIE23}. In the example in Fig.~\ref{fig:nonequilibrium} it is due to a non-linear, multiplicative dependence on the control parameter.
It can also happen due to non-smooth saddle-nodes (Stommel's model \cite{STO61,LOH21b}), as well as for tipping via homoclinic and subcritical Hopf bifurcations (Fig.~\ref{fig:bifurcations}a,b) and thus in many AMOC box models \cite{SCO99,TIT02,TIT02a,ALK19} and Welander's model of the convective feedback (see Appendix~\ref{appendix:welander}) \cite{WEL82,CES96,ABS04}.
How prevalent it is in state-of-the-art models remains to be seen. It has been shown for the AMOC in ocean-only and coupled models \cite{STO97,LOH21,VWE26b} - with recent studies providing mechanistic insights \cite{HAN25,VWE26b} - and there is evidence for it in ice sheet collapses \cite{SWI26b,FEL26}.

In the realm of climate models, which cannot be initialized in the far past (with past forcing) due to the high computational cost, a simulation can even be out-of-equilibrium with the pullback attractor (B4). This might be referred to as ``triple'' non-equilibrium. In spinup simulations, models are usually simulated with fixed climatological forcing until they reach a quasi-steady state, and this is then used as initial condition for experiments (e.g. starting in the industrial period) with varying natural and anthropogenic forcings. This is necessary such that there are no spurious relaxation trends that bias the projected response due to the forcing. However, the target here is a NESS of the model and not the historic pullback attractor.
Thus, simulations can fail to track the pullback attractor, and the model-predicted TP can depend on how far away the initial conditions were from the pullback attractor at the time of simulation start \cite{LOH24d} (see blue trajectories in Fig.~\ref{fig:nonequilibrium}b).
The issue is also known for ice sheet models' predictions of future sea level rise, which ideally need to be initialized with simulations of the entire last glacial period and estimated surface temperature forcing \cite{ASC13}.

EWS rely on being close to the NESS up until shortly before the bifurcation. But for realistic forcing rates, the slowing dynamics in the critical d.o.f may only be detectable long after the point of no return (in the fixed-parameter system) has been crossed \cite{BOL21}.
Figure~\ref{fig:nonequilibrium}c,d shows this for the saddle-node normal form system. For an ensemble of realizations (in gray) one observes an increase in autocorrelation towards 1 and a near-divergence of variance just before tipping ($x<0$). This warning only comes long after the bifurcation.
However, the situation is even worse in reality, since EWS need to be estimated not from ensembles but individual time series.
Here, the underlying trend from the moving NESS needs to be estimated, after which the changing statistical properties can be estimated, e.g., in a sliding window. Even with known trend, this yields far lower inferred increases in variability (dotted lines in Fig.~\ref{fig:nonequilibrium}d).
If it is unknown, it can be removed by a filter, or within each window by linear/spline detrending (dot-dashed lines in Fig.~\ref{fig:nonequilibrium}d), yielding an even lower signal as more low-frequency variability is removed. To avoid sliding windows, one could perform a maximum-likelihood fit or Bayesian inference of a time-dependent stochastic process \cite{DIT23,HAL25},
but this hides the large uncertainty of assuming a particular model (a normal form only holds close to the bifurcation) and time dependence \cite{LOH26}.
Similarly, for non-autonomous instabilities, it is uncertain whether the quasi-equilibrium indicators hold \cite{LOH21b}, and whether relevant observables for EWS (see next section) remain the same. 


In the non-autonomous case, there is a range of forcing rates and initial conditions where the system may either tip or not \cite{ALK18,ASH21}. This is true for rate-induced tipping, non-monotonic forcings across a bifurcation (overshoot), or when neither of this is the case but when allowing for initial conditions outside of the (pullback) attractor (triple non-equilibrium). 
Moreover, for chaotic systems, due to chaotic mixing during the parameter shift or fractal basin boundaries this dependence of tipping becomes sensitive, i.e., there is a loss of predictability \cite{LOH24d} where very nearby simulations (in terms of initial conditions or forcing trajectory) can have a different outcome (tipping or not), and at the same time would show the same EWS (if these exist), leading to false positives.
Climate models have demonstrated that such a splitting of the ensemble with respect to the qualitative outcome at the end of a simulation can occur \cite{KAS19,LOH21,ROM23,LOH24d,OH25,BOE26}.

Finally, there may be far-from-equilibrium dynamics (B5), which in boundary cases of non-autonomous instabilities are due to a chaotic saddle (edge state) \cite{LOH21}, but may also occur after crossing a crisis in the autonomous system due to ghost states \cite{BOE26}. This leads to long transients with lifetimes that are essentially unpredictable. It further limits the predictability of tipping, revealing another caveat of local stability analysis, but also gives rise to a window of opportunity (see also the 'dynamic tipping window' in \cite{ASH25b}) to return to the base attractor when reversing the forcing.

\subsection{Observing, early-warning, and attribution}
\label{sec:ews}

The theoretic challenges highlighted in the previous two sections all have implications for the possibility to detect TPs from observational data, see right column of Fig.~\ref{fig:schematic}. Note that these are in addition to various statistical challenges with EWS \cite{RIE25}, such as multiple testing and spatial autocorrelation, or changes in data quality, availability, homogeneity and stationarity.

Because of high-dimensional dynamics not every scalar observable will display EWS \cite{BOE13,LOH25}. This is because the CSD happens in a low-dimensional sub-space (indicated by the $x_1 - x_2$-plane in Fig.~\ref{fig:csd}), which in the case of a saddle-node bifurcation is related to the center manifold. Many other degrees of freedom may not project significantly onto it until arbitrarily close to the TP.
Even if an observable is a plausible and established summary statistic of the system state, and can clearly distinguish the present-day from an alternative state, it may not show increased fluctuations (EWS). For instance, the AMOC's volume transport does not show EWS before an AMOC collapse in several models \cite{VWE24,LOH25}. Thus, it is likely not robust to look for the best ``proxies'' of a given system (such as for the AMOC strength \cite{BOE21a,DIT23}) and use this as scalar observable for EWS in data.

Similarly, there can be increases in variability in various d.o.f that are not the result of CSD \cite{RIE25}, but are due to smooth changes in oscillatory modes. Indeed, recent studies analyzing a simulated AMOC collapse search for regions and variables with high increase in variability to use as EWS \cite{SMO25,PAT26}, but find equally many regions/variables with a significant {\it decrease}, also in regions of plausible relevance to the AMOC. In addition, there is often a large disagreement between the evolution of variance and autocorrelation \cite{BOU14,SMO25}, casting doubt on whether the signals are related to CSD. This corroborates that if one does not know the correct quantity, observing a statistical EWS in a given observable does not by itself provide sufficient evidence for an impending TP.
The observational record (e.g. of sea surface temperature and salinity \cite{BEN23}, and of precipitation \cite{BOC23}) also features both increases and decreases in variability for closeby regions, making current evidence for EWS highly uncertain.

Hence, if a particular statistical signal should be attributed as sign of an impending TP, it needs to have a distinct relation to the tipping mechanism.
Since the largest fluctuations can be expected along a path towards the edge state, it has been proposed to estimate the edge state in (sufficiently realistic) models of the system in question, and deduce observables where it stands out most \cite{LOH25}. Related approaches to find the d.o.f with largest fluctuations in models include finding reaction coordinates \cite{ZAG24}, rare-event algorithms \cite{CIN24}, and instanton computations \cite{SOO25}. It was also proposed to find physics-based EWS by choosing observables that have a direct connection to the suspected underlying positive feedback \cite{VWE25}.

Edge states or instantons are very difficult to obtain and may be model-dependent. Physics-based observables may be ambiguous, depending on how well the responsible feedback can be defined. An alternative are methods that directly target the critical d.o.f. from high-dimensional data sets. The critical d.o.f before tipping should be related to the slowest Kolmogorov mode \cite{LUC24,LOH25b}, assuming a low-noise limit where noise-induced transitions are not observed. 
Data-driven approaches for estimating this mode include approximations of the stochastic Koopman operator \cite{GUT22}, for instance using (kernelized) dynamic mode decomposition \cite{ZAG26, LUC26b}, approximations of the transfer operator via Ulam's method and reduced state space Markov chain modeling \cite{TAN18,LUC26}, 
as well as estimation of the generator (Kolmogorov operator) via diffusion maps \cite{LOH25b}.
These approaches need to be further generalized to address the non-equilibrium issues raised in the previous section (C2,C3).


\begin{figure}
\includegraphics[width=0.8\textwidth]{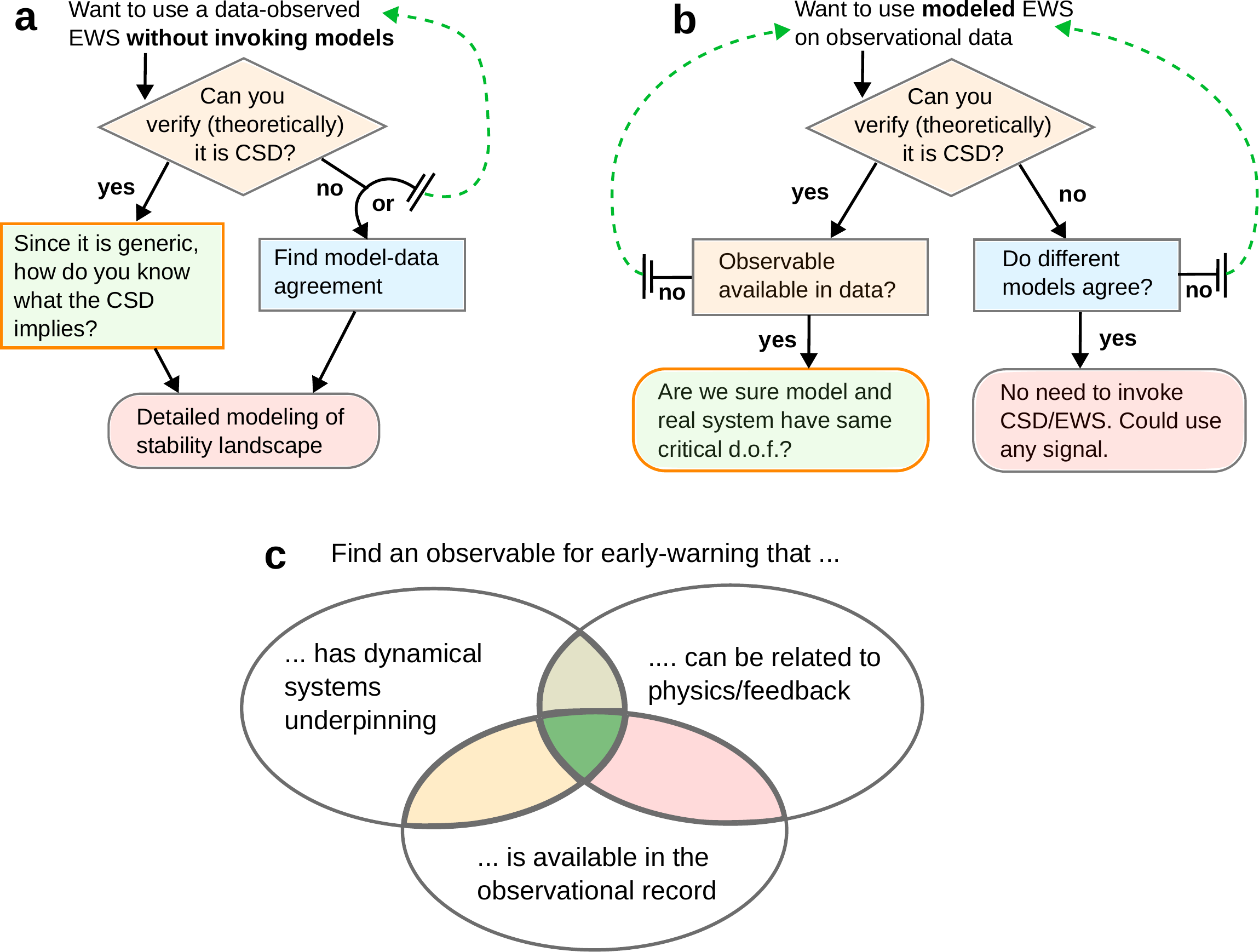}
\caption{\label{fig:catch22}
{\bf a,b} The Catch-22 of early-warning signals (EWS) either when observed in data ({\bf a}) or model simulations ({\bf b}), as further explained in the main text. {\bf c} Proposition for best-practice in the search of observables that can be used for EWS.
}
\end{figure}

Even assuming that CSD exists in the most general case (may not be the case for strong non-equilibrium), 
in practice statistical EWS are not as ``generic'' as hoped and their usage may lead to a situation reminiscent of a ``catch-22''.
That is, by trying to rigorously show that a statistical EWS is indeed due to CSD, one may implicitly go against the intention of using EWS in the first place, as explained in the following and Fig.~\ref{fig:catch22}a,b.

After identifying a statistical EWS in either data or model simulations, the first central question is whether one can verify that the signal is actually due to CSD, using the abovementioned methods while taking into account statistical properties of data and assumptions on the noise process. If this can be done for data (Fig.~\ref{fig:catch22}a), the catch is that since CSD and the EWS are generic, one does not know what they imply, i.e., a major or only minor state transition. To find out requires detailed modeling of the system's stability landscape. This is against the original intention, which was to leverage generic properties of dynamical systems to warn of a TP without the need of a realistic model (if even available). If one cannot verify CSD, then one either has to start over and find another candidate EWS (decreasing statistical significance), or as a compromise try to obtain a model-data agreement, which again conflicts with the model-free intention.

If for a modeled statistical EWS (Fig.~\ref{fig:catch22}b) it can be verified it is CSD (operator methods, edge states, etc.), the question is whether the corresponding observables are available in existing data of sufficient length. If yes, one is in a very good position and could test whether the signal is indeed observed in data, but even then there is a catch since we cannot necessarily be sure that the critical degree of freedom is the same in model and real-world system. For instance, modeled modes of variability or convection in the turbulent ocean are still limited by model resolution.
If CSD cannot be verified, one could start over, or compromise and establish that the signal is common to most models. This would strengthen its relevance to predicting the specific TP in question, but it remains just a statistical signal, and there is no need to invoke CSD. That is, instead of signals consistent with the CSD idea (increases in fluctuations and correlations), any statistical signal (common to multiple models) could be used, which may have stronger power than the one originally proposed.


This purposely pessimistic view is not meant to convey that EWS cannot be used
as one of several lines of evidence for TPs. Current best-practice for the analysis of EWS may be to look for a self-consistent agreement of data, theory, physical understanding and numerical modeling, as illustrated in Fig.~\ref{fig:catch22}c. One should derive (from realistic models) critical observables with a dynamical systems underpinning, which are in good agreement with what is expected from the relevant feedback processes, and which are available in the observational record with sufficient statistical properties. If then observational data show EWS, one could consider this evidence for an approaching TP. In addition, it would be good to show in realistic models that there is indeed a clear, singular large-scale TP that is not riddled by higher multistability and intermediate TPs.
In principle, it may be that such multistability a) disappears due to merging of closeby coexisting states by small-scale dynamics (acting like bounded noise), b) is an artifact due to parameterizations and discretization in lower-resolution models, or c) becomes so fine-scaled it effectively appears continuous as more scales and processes are included.

\section{Possible directions for research}
\label{sec:directions}

\subsection{Analysis of climate models}
More resources should be invested in analyzing the global stability landscape in complex models of various geophysical systems. This will help understand what a climate TP really is from a mathematical viewpoint in the quasi-equilibrium setting, with implications also for other complex systems. Besides ongoing efforts to quantitatively compare critical forcing thresholds, it would be insightful to compare the dynamical systems features, such as the characteristics of attractor {\it crises}, between several models and across Earth system components.

Researching the models' basins of attraction would help to understand the prevalence of regions with tightly folded or fractal basins. 
Besides chaotic dynamics, such regions lead to ensemble splitting, as observed in climate models \cite{KAS19,LOH21,ROM23,LOH24d,OH25,BOE26}. The larger the extent of such regions, the larger the ``fuzziness'' of the TP (loss of predictability) when varying initial conditions and forcing scenarios. In addition, such research would help to estimate how widespread basin instability and rate-induced tipping are.

A comparison of edge states in different models \cite{LOH24b,BOE26,MOI26} might uncover useful agreements in tipping mechanism and possible EWS, even if the forcing thresholds are very different. A comparison of Koopman/Kolmogorov operators and modes across models would help further understand the nature of the slowest relaxation modes, whether these are obtainable in observational data, and whether they can be used to predict the near-term response to different climate forcings.

Coming research will also clarify whether high multistability seen in individual models \cite{MAR21,RAG22,LOH24,ADL24} survives when increasing model complexity and resolution, or whether many previously co-existing attractors merge and the stability landscape again simplifies to leave only the possibility of a few, large-scale TPs. In models of intermediate complexity, it would be interesting to compute a ``coupled'' bifurcation diagram of, e.g., the AMOC and the Greenland/West Antarctic ice sheets as function of $CO_2$ levels. This combines ice sheet meltwater forcing and global warming, often modeled as separate control parameters, as well as additional feedbacks between AMOC and the ice sheets. If the AMOC and ice sheets have intermediate states in isolation, it will be interesting to see whether the multistability of the coupled system becomes more complex, or whether it is overall stabilized \cite{POP25}.

The computational cost for such suggestions is very large. Hopefully the next generation of GPU-accelerated models will help \cite{HAE21,YAT26}, but ideally, to make non-incremental progress, models need to be probed in ways that are more efficient compared to forward simulations. One approach are fully implicit models, with which one can compute bifurcation diagrams via numerical continuation  \cite{THI26}, which are however restricted to equilibria.
Some questions, such as the sensitivity of stable states to different parameters, may be more efficiently answered with adjoint models \cite{SEV17}.
Similarly, fully differentiable models \cite{MEU25,DAV26,MOS26} would allow for gradient-based optimization of parameters and model trajectories, such that
transition pathways to different regimes may be found.
This may also be done efficiently using large-deviation theory \cite{SOO25} and rare-event algorithms \cite{CIN24}, where it would be interesting to implement self-adapting score functions (see e.g. \cite{ESC26}) such that an ensemble of simulations would explore phase space more globally, instead of only one particular rarest trajectory.
This may be used to establish that there are TPs and multistability in computationally very expensive models.

\subsection{Analysis of past and present transitions in data}
More work is warranted on finding and scrutinizing viable analogues for past TPs, even if these are on a smaller, regional scale. Preferably, they should not be in the ``geological past'', because proxy observations are uncertain and non-stationary. Besides verifying EWS, the quest for past small-scale tipping is important, since it would corroborate the proposal of fragmented, intermediate tipping \cite{RIE21,BAS22,LOH24} and multiscale multistability \cite{MAR21}.
Candidates may be found in the ocean circulation, for instance transitions from deep to intermediate convection in the Greenland sea \cite{STR24} or in boundary currents \cite{SCH01,VWE26}.

From a methodological point of view, both past and present climate observations are sparse and this may be in conflict with ideal observables for EWS. Hence, theory and methods should be developed that can assess CSD from partial observations, and in particular from shorter-timescale variables if the critical d.o.f. evolves on time scales longer than the observational horizon. An example would be to assess CSD of the AMOC from surface ocean observations if the critical d.o.f. lies in the deep ocean \cite{LOH25b}.
CSD might be seen in higher (i.e. faster) eigenmodes, but one may need to have data even closer to the TP to succeed. This may increase the risk of false positives as it may become more uncertain that tipping will encompass the entire system as initially suspected. 
One approach could be a ``time-for-space'' substitution, i.e., an attractor reconstruction based on Takens embedding, which can incorporated into an analysis of Koopman/Kolmogorov modes \cite{DAS19,FRO21}.
A drawback is that it requires long time series and may limit physical interpretability.

Note that this is in some sense opposite to ``space-for-time'' substitutions for EWS \cite{GUT09,CHE12}. Measuring spatial instead of temporal correlation can increase early-warning skill (for given length of time series) since no detrending  and windowing is needed, given that spatial locations are sufficiently independent (weakly coupled) \cite{CLA26}. But when relevant observations and spatial dimensions are missing, this may not be possible.

\subsection{The role of the noise process}

As EWS rely on stochastic fluctuations, it has been questioned whether Gaussian white noise is an appropriate assumption.
Several extensions for EWS have been proposed, including for multiplicative \cite{MOR24,LOH25b}, red \cite{BOE22,MOR24b,BER26}, non-markovian \cite{KUE22},
correlated (non-diagonal) \cite{MOR23}, non-Gaussian ($\alpha$-stable) \cite{LAY25},
degenerate (e.g. for the ocean forced only at surface and boundaries) \cite{BER24,BER26},
and nonstationary noise \cite{BOE22,MOR24}. 
These studies show that, while CSD as such largely remains intact, in many cases standard EWS are less reliable, decrease statistical significance, and increase the possibility for false positives or negatives, thus requiring a variety of different statistical techniques.
To come to a conclusion in regards to such techniques, it should be scrutinized what type of noise model is appropriate for different components of the Earth system.
This could perhaps be done by dedicated model simulations at varying (high) resolutions and a comparison to data.
In turbulence-resolving simulations it would also be interesting to clarify the relation of EWS and changes in scaling of turbulent fluctuations before transitions, as well as different possible limiting cases for noise processes that arise from fast chaotic geophysical dynamics.

These investigations may also motivate whether a low-noise assumption holds, which is relevant as otherwise noise-induced transitions could be the norm before EWS close to a bifurcation are detectable.
This is related to constraining the probability of noise-induced transitions from internal variability \cite{CIN24,CIN26}, or other processes such as volcanic eruptions.
From observations as well as simulations forced with anthropogenic emissions it may also be scrutinized whether whatever is considered the ``noise'' is stationary in strength and correlation.
Given climate change also affects atmospheric variability 
this deserves strong justification, as it crucially needs to be taken into account for EWS. 


\subsection{EWS theory out of equilibrium}

Important theoretic advances could be made to clarify the existence of EWS in Earth system components that are forced rapidly away from their NESS (B2), and possibly display non-autonomous transitions (B3). This would be important especially if EWS are used to evaluate whether a TP is already underway, or when precisely it can be expected to happen. One direction would be an extension of the above-mentioned operator-theoretic methods to non-autonomous dynamical systems.
Another is the utilization of modern ideas and methods in non-equilibrium thermodynamics and statistical physics.
For instance, it has been proposed to measure time-irreversibility in time series via asymmetry of correlation functions, serving as quantifier of the extent of detailed balance breaking, which has been purported to change prior to a bifurcation \cite{YAN23,XU23,KOO26}. This seems to yield signals in some systems, but requires a better theoretical justification, especially given that the Earth system is likely already out-of-equilibrium in several ways also when far from a bifurcation.
A deeper engagement with recent developments in stochastic thermodynamics \cite{SEI26}, and how they may be applicable in the climate context, could be fruitful.

\section{Conclusion}

Sudden, nonlinear changes in parts of the climate system are now simulated with sophisticated models, sometimes even under moderate emission scenarios. While this indicates the risk of irreversible TPs, it is uncertain whether the simulated transitions are due to true multistability and bifurcations that would allow their data-driven prediction from generic precursors (EWS).
It is still doubtful that we can anticipate large-scale TPs and with useful certainty distinguish them from gradual, reversible changes or minor reorganizations of the system. This is due to limits in our current knowledge on general systems out of equilibrium, and on the prevalence and nature of regime transitions in the physical sub-systems of the climate in particular. To make progress, the dynamical properties of the stable and unstable steady states of more climate models, and more generally their global stability landscape, needs to be analyzed in detail.
On the observational side, if a signal is to be attributed to an impending TP, it should come from a non-trivial, information-rich fingerprint that can distinguish an actual TP from reversible changes, even when far from equilibrium. This may be achievable by combining new approaches based in dynamical systems and statistical mechanics with expert insights into the physical processes. Machine learning may soon be rolled out for predicting future climate. This would happen in the precarious situation where observational training data is very limited compared to the extent and number of physically possible dynamical regimes.
Hence, now is the right time to learn as much as possible about complex systems and their potential multistability and bifurcations.





\bibliography{refs} 

\appendix
\setcounter{table}{0}
\renewcommand{\thetable}{A\arabic{table}}

\setcounter{figure}{0}
\renewcommand{\thefigure}{A\arabic{figure}}

\section{On rate-induced tipping and subcritical Hopf in Welander's model}
\label{appendix:welander}
The model introduced by Welander \cite{WEL82} to describe deep convection features a Hopf bifurcation that lead to oscillations between a deep convective and a stratified state. There is a parameter region where this Hopf bifurcation becomes subcritical, whereby rate-induced tipping is possible via the mechanism discussed in Sec.~\ref{sec:global} and Fig.~\ref{fig:bifurcations}a,b. The model dynamics switch between a convective and non-convective state around a vertical density gradient threshold $\epsilon$. This switching can be described by a sigmoid function $\nu_{\epsilon}$, as in \cite{ABS04}, resulting in the following the governing equations
\begin{align}
\dot{x} &= 1 - x - \nu_{\epsilon} (x,y) x\\
\dot{y} &= \mu - \nu_{\epsilon} (x,y) y
\end{align}
with the function $\nu_{\epsilon}(x,y) = \nu_1 + \frac{1}{2} (\nu_2 - \nu_1) [1 + \tanh (\sigma(y - x + \epsilon))]$.
The non-dimensional state variables $x$ and $y$ represent the vertical temperature and salinity gradients, respectively. The constants $\nu_1 = 0.1$ and $\nu_2=5.0$ set the size of the jump in vertical diffusivity, and $\sigma=8$ is the steepness of the sigmoidal shift in dynamics. The control parameter $\mu$ is proportional to the freshwater forcing at the surface.

\begin{figure}
\includegraphics[width=0.99\textwidth]{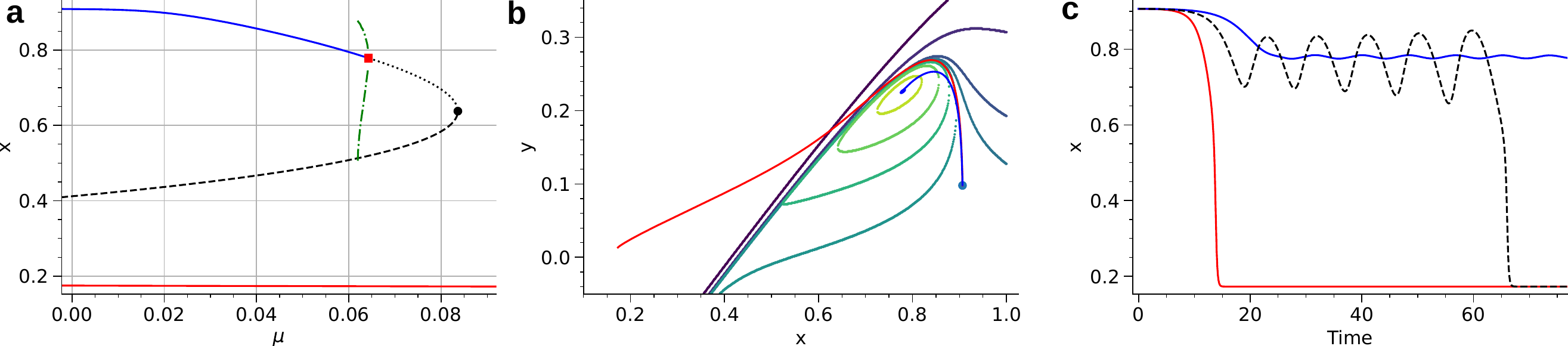}
\caption{\label{fig:welander}
{\bf a} Bifurcation diagram of the Welander model with parameter $\epsilon = 0.35$. Stable fixed points are solid lines, with the convective state in blue. Unstable fixed points are the dotted and dashed lines, and the dash-dotted line is the unstable periodic orbit (indicated by its maximum and minimum $x$ values). Fold and Hopf bifurcations are dot and square, respectively.
{\bf b} Phase portrait of the model, where the colored thick lines are the basins of attractions at different values of $\mu$, starting at $\mu = 0.1$ (dark colors) and ending at $\mu = 0.064$ (light colors), which is shortly before the subcritical Hopf bifurcation. The red (blue) lines is a trajectory starting at the convective state at $\mu=0.1$ (blue dot), whereafter $\mu$ is ramped linearly to a final value of $\mu=0.064$ within 10 (25) time units, after which $\mu$ is kept constant.
{\bf c} Corresponding time series, with an additional simulation at an intermediate duration of the ramping of 17 time units (dashed line).
}
\end{figure}

Figure~\ref{fig:welander}a shows a bifurcation diagram with respect to $\mu$. An unstable limit cycle exists between a homoclinic bifurcation (where it touches the saddle point at the dashed line) and the subcritical Hopf bifurcation (red square). Leading up to the homoclinic bifurcation, the basin of attraction of the convective state (blue line) shrinks rapidly and becomes finite after the bifurcation (Fig.~\ref{fig:welander}b). Thus, starting with an initial condition in the convective state at low $\mu$ and thereafter increasing $\mu$ over time can lead to a trajectory that falls outside the basin (at the instantaneous, fixed parameter value) of the convective state before the subcritical Hopf is reached.

If the parameter shift is fast enough - and even if it stops before the Hopf bifurcation - the trajectory tips to the non-convective state, as shown in Fig.~\ref{fig:welander}b,c. Close to the limiting case of a critical rate of the parameter shift, the trajectory approaches the unstable periodic orbit and shows some oscillations before a transition to either fixed point, as shown in the dashed time series in Fig.~\ref{fig:welander}c.

This subcritical scenario disappears and becomes supercritical for smaller $\epsilon$. It is not argued here which scenario is more realistic in the real ocean.

\end{document}